# Volumetric dose extension for isodose tuning

Lin Ma[1], Mingli Chen[1], Xuejun Gu[†1] and Weiguo Lu[†1]

[1]*Medical Artificial Intelligence and Automation Laboratory, Department of Radiation Oncology, University of Texas Southwestern Medical Center, 2280 Inwood Rd, Dallas, TX 75390, USA*

[†]Corresponding Author: Xuejun.Gu@UTSouthwestern.edu  Weiguo.Lu@UTSouthwestern.edu

**Abstract**

Purpose: To develop a method that can extend dose from two isodose surfaces (isosurfaces) to the entire patient volume, and to demonstrate its application in radiotherapy plan isodose tuning.

Methods: We hypothesized that volumetric dose distribution can be extended from two isosurfaces—the 100% isosurface and a reference isosurface—with the distances to these two surfaces ($L_{100}$ and $L_{ref}$) as extension variables. The extension function is modeled by a three-dimensional lookup table (LUT), where voxel dose values from clinical plans are binned by three indexes: $L_{100}$, $L_{ref}$ and $D_{ref}$ (reference dose level). The mean and standard deviation of voxel doses in each bin are calculated and stored in LUT. Volumetric dose extension is performed voxelwisely by indexing the LUT with the $L_{100}$, $L_{ref}$ and $D_{ref}$ of each query voxel. The mean dose stored in the corresponding bin is filled into the query voxel as extended dose, and the standard deviation be filled voxelwisely as the uncertainty of extension result. We applied dose extension in isodose tuning, which aims to tune volumetric dose distribution by isosurface dragging. We adopted extended dose as an approximate dose estimation, and combined it with dose correction strategy to achieve accurate dose tuning.

Results: We collected 32 post-operative prostate volumetric modulated arc therapy (VMAT) cases and built the LUT and its associated uncertainties from the doses of 27 cases. The dose extension method was tested on five cases, whose dose distributions were defined as ground truth (GT). We extended the doses from 100% and 50% GT isosurfaces to the entire volume, and evaluated the accuracy of extended doses. The 5mm/5% gamma passing rate (GPR) of extended doses are 92.0%. The mean error is 4.5%, which is consistent to the uncertainty estimated by LUT. The dose difference in 90.5% of voxels is within two sigma and 97.5% in three sigma. The calculation time is less than two seconds. To simulate plan isodose tuning, we optimized a dose with less sparing on rectum (than GT dose) and defined it as a "base dose"—the dose awaiting isosurface dragging. In front-end, the simulated isodose tuning is conducted as such that the base dose was given to plan tuner, and its 50% isosurface would be dragged to the desired position (position of 50% isosurface in GT dose). In back-end, the output of isodose tuning is obtained by 1) extending dose from the desired isosurfaces and viewed the extended dose as an approximate dose, 2) obtaining a correction map from the base dose, and 3) applying the correction map to the extended dose. The accuracy of output—extended dose with correction—was 97.2% in GPR (3mm/3%) and less than 1% in mean dose difference. The total calculation time is less than two seconds, which allows for interactive isodose tuning.

Conclusions: We developed a dose extension method that generates volumetric dose distribution from two surfaces. The application of dose extension is in interactive isodose tuning. The distance-based LUT fashion and correction strategy guarantee the computation efficiency and accuracy in isodose tuning.

**Symbols**

| | |
|---|---|
| $\vec{x}$ | 3D point described by Cartesian Coordinates. $\vec{x}_q$ is query point and $\vec{x}_{\text{ref}}$ is reference point. |
| $S$ | Surface embedded in 3D Euclidean space. $S_{100\%}$ is 100% isodose surface. $S_{\text{ref}}$ is reference surface. |
| $V$ | The volume of patient in 3D Euclidean space. |
| $D$ | 3D dose distribution defined in $V$. |
| $\widetilde{D}$ | Approximate dose distribution. Extended dose is used as approximate dose in this study. |
| $\ddot{D}$ | Actual dose distribution, which is associated to an existing plan. |
| $L_{100}$ | A scalar field (distance map) that associates the value of $\mathcal{L}_{100}$ to every query point $\vec{x}_q$ in $V$. |
| $L_{\text{ref}}$ | A scalar field (distance map) that associates the value of $\mathcal{L}_{\text{ref}}$ to every query point $\vec{x}_q$ in $V$. |
| $D_{\text{ref}}$ | A scalar field (dose map) that associates the value of reference dose to every query point $\vec{x}_q$ in $V$. |
| $\mathcal{F}$ | A function that extend dose from two surfaces ($S_{100\%}$ and $S_{\text{ref}}$) to volume $V$. |
| $\mathcal{L}_{100}$ | A function that measures the distance between a point and 100% isodose surface. |
| $\mathcal{L}_{\text{ref}}$ | A function that measures the distance between a point and reference surface. |

## 1   Introduction

Radiation dose distribution is determined from delivery parameters by dose calculation algorithm[1, 2], which models the transportation of particles (photons, electrons, …) and energy deposition in patient volume. Inverse planning[3], an optimization procedure, is used to obtain the delivery parameters corresponding to the desired dose distribution in clinic. To maximize the killing on tumor while minimize the harm to normal tissue, the optimization objectives enforce the optimized dose to have high coverage of planning target volume (PTV), fast dose falloff outside PTV and sparing of organs at risk (OARs). Multiple studies have shown that optimized dose distributions are heavily influenced by and correlated to the geometrical relationship between PTVs and OARs. These studies can be categorized into to two groups: explicit distance-based methods and implicit distance-based methods. Explicit distance-based methods modelled dose as an explicit function of distance to PTVs. The works[4, 5] of knowledge-based planning belong to this category. They calculated the histogram of distance to PTV and used principle component analysis as the function to obtain dose from distance. Another work[6] expanded the dose distribution from PTV to patient volume using an explicit function of distance to PTV. The function was composed of multiple terms with different falloff speeds, which modeled physics effects such as penumbra-driven dose gradient and cross-fire effect. Implicit distance-based methods directly predicted dose distribution from the distribution of regions of interest (ROIs) without using distance as an explicit variable. The implicit methods were realized by deep learning (DL) networks that were trained to directly map ROI distribution to dose distribution, and have been applied in many disease sites such as prostate[7], head and neck[8, 9], lung[10] and esophagus[11]. Most DL networks achieve a map from the ROI distribution to a single population-based dose distribution, but lack the flexibility to tune towards specific target/OAR tradeoff. Only few works attempt to add additional information in input[12, 13] or output[14] to make the predicted dose tunable, but none of them is flexible enough to approach isodose tuning.

We presented a novel yet simple volumetric dose extension method in this paper. This method is an explicit distance-based method, which analytically extends dose from two isodose surfaces (isosurfaces) to the entire patient volume, with distances to the two surfaces as extension variables. We employed the 100% isosurface $\boldsymbol{S}_{100\%}$ and a reference surface $\boldsymbol{S}_{\text{ref}}$ in this research. The volumetric extension operation effectively extend dose to the entire dose volume space (figure 1)—extrapolation towards outside of $\boldsymbol{S}_{\text{ref}}$, extrapolation towards inside of $\boldsymbol{S}_{100\%}$ and interpolation in between two surfaces. The extension function is represented by a lookup table (LUT, figure 2), whose values are derived from clinical doses. It only requires a small number of plans to build the LUT and the extension is performed by table lookup. Therefore, the implementation and application of this dose extension method are easy and fast. Moreover, the introduction of reference isosurface distinguishes this method from the abovementioned methods which can only generate a single population-based dose distribution. With 100% isosurface $\boldsymbol{S}_{100\%}$ fixed, the information of dose falloffs along different directions is described by the position of reference isosurface $\boldsymbol{S}_{\text{ref}}$ (figure 1). The dose extended from such a combination of two isosurfaces is very flexible, which can be tuned by isosurface dragging.

The rest of paper is organized as follows. In the Methods and Materials section, we describe the dose extension method and its application in isodose tuning. Section 2.1 articulates the formulation and LUT representation of extension function. Section 2.2 details the application of dose extension in the scenario of isodose tuning. Sections 2.3 presents the two experiments designed for dose extension itself and its application in isodose tuning. In the Results section, we firstly report the extension function—LUT—learned from clinical doses in section 3.1. Then we reported the accuracy of dose extension and isodose tuning in section 3.2 and 3.3, respectively. In the Discussion section, we summarize results and show directions for future research. For simplicity without loss of generality, all dose values are normalized to the prescription throughout this paper unless specified differently.

## 2 Methods and Materials
### 2.1 Volumetric dose extension
#### 2.1.1 Formulate dose extension

We hypothesize that the dose distribution in a volume can be extended from the dose distributions on two surfaces (figure 1). In this study, we focus on extending dose from the 100% prescription dose isosurface $\boldsymbol{S}_{100\%}$ and a reference surface $\boldsymbol{S}_{\text{ref}}$.

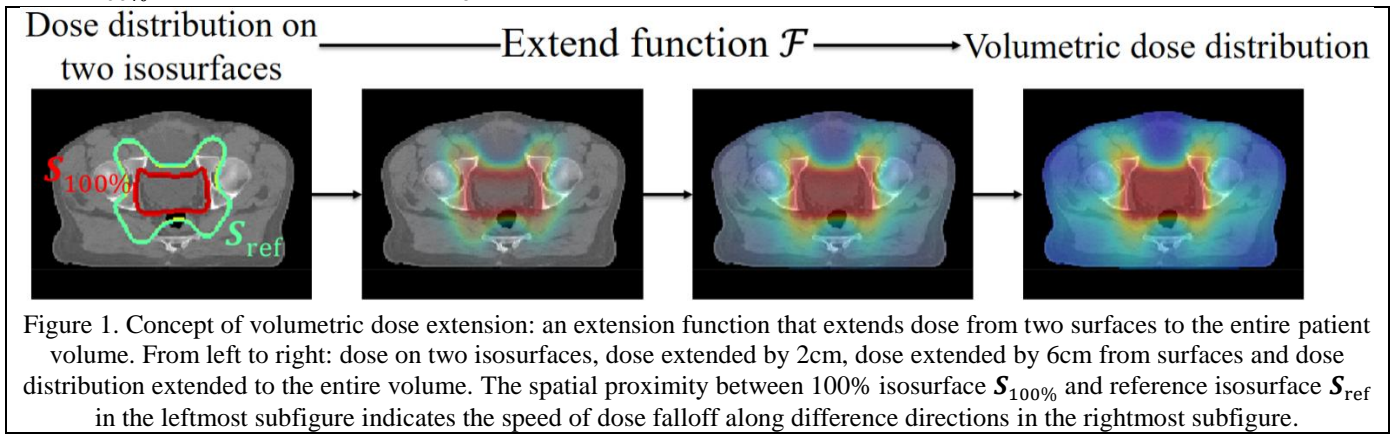

Figure 1. Concept of volumetric dose extension: an extension function that extends dose from two surfaces to the entire patient volume. From left to right: dose on two isosurfaces, dose extended by 2cm, dose extended by 6cm from surfaces and dose distribution extended to the entire volume. The spatial proximity between 100% isosurface $\boldsymbol{S}_{100\%}$ and reference isosurface $\boldsymbol{S}_{\text{ref}}$ in the leftmost subfigure indicates the speed of dose falloff along difference directions in the rightmost subfigure.

Equation (1) shows a general form of extension function $\mathcal{F}$, which outputs the dose in a query point (voxel). The inputs of $\mathcal{F}$ are the geometrical positions of query point $\vec{x}_q$ and two surfaces, and the dose distributions on the two surfaces. As $D[\![S_{100\%}]\!] \equiv 1$, only the dose distribution on reference surface $D[\![S_{\text{ref}}]\!]$ is in the variables of $\mathcal{F}$.

$$D = D[\![\vec{x}_q]\!] = \mathcal{F}(\vec{x}_q, S_{100\%}, S_{\text{ref}}, D[\![S_{\text{ref}}]\!]) \tag{1}$$

We further modelled the dose of query point as a function of its distances to the two surfaces. Therefore, the explicit form (2) of extension function $\mathcal{F}$ has three variables: $L_{100}$, $L_{\text{ref}}$ and $D[\![S_{\text{ref}}]\!]$. $L_{100}$ is the distance from point $\vec{x}_q$ to surface $S_{100\%}$ measured by function $\mathcal{L}_{100}$, while $L_{\text{ref}}$ is the distance from point $\vec{x}_q$ to surface $S_{\text{ref}}$ measured by another function $\mathcal{L}_{\text{ref}}$.

$$D = D[\![\vec{x}_q]\!] = \mathcal{F}(\mathcal{L}_{100}(\vec{x}_q, S_{100\%}), \mathcal{L}_{\text{ref}}(\vec{x}_q, S_{\text{ref}}), D[\![S_{\text{ref}}]\!]) \tag{2}$$

*2.1.2  Definition of distances*

The function $\mathcal{L}_{100}$ is defined in (3) as the signed Euclidean distance from query point $\vec{x}_q$ to the closest point on $S_{100\%}$. The distance is negative for a query point inside the 100% isosurface. $L_{100}$ is the scalar field that associates value of $\mathcal{L}_{100}$ to every query point $\vec{x}_q$ in patient volume $V$. An example of $L_{100}$ is in supplementary material (A).

$$L_{100}: \mathcal{L}_{100}(\vec{x}_q, S_{100\%}) = \min_{\vec{x}' \in S} ||\vec{x}_q - \vec{x}'||_2^2 \tag{3}$$

Concerning the definition of $\mathcal{L}_{\text{ref}}$, we firstly calculate the geometric center of target volume $\vec{x}_c$ and locate the reference point $\vec{x}_{\text{ref}}$ by (4), where $\vartheta(\vec{x})$ evaluates the orientation of vector $\vec{x}$. Reference point is the point on $S_{\text{ref}}$ that is collinear with $\vec{x}_q$ and $\vec{x}_c$. The position of reference point $\vec{x}_{\text{ref}}$ changes continuously to the position of query point $\vec{x}_q$.

$$\vec{x}_{\text{ref}} = \underset{\vec{x}' \in S_{\text{ref}}}{\arg\min} |\vartheta(\vec{x}_q - \vec{x}_c) - \vartheta(\vec{x}' - \vec{x}_c)| \tag{4}$$

The function $\mathcal{L}_{\text{ref}}$ is defined in (5) as the signed Euclidean distance from query point to reference point. The distance is negative for a query point inside the reference surface. $L_{\text{ref}}$ is the scalar field that associates value of $\mathcal{L}_{\text{ref}}$ to every query point $\vec{x}_q$ in $V$. An example of $L_{\text{ref}}$ is in supplementary material (B).

$$L_{\text{ref}}: \mathcal{L}_{\text{ref}}(\vec{x}_q, S_{\text{ref}}) = ||\vec{x}_q - \vec{x}_{\text{ref}}||_2^2 \tag{5}$$

We further introduce the notation $D_{\text{ref}}$ in (6), which is a scalar field that associates to every query point $\vec{x}_q$ in $V$ the dose of its reference point $D[\![\vec{x}_{\text{ref}}]\!]$. As the correspondence from $\vec{x}_q$ to $\vec{x}_{\text{ref}}$ is continuous, the scalar field $D_{\text{ref}}$ is continuous. $D_{\text{ref}} \equiv z\%$ when the reference surface is z% isosurface. Examples of $D_{\text{ref}}$ are in supplementary material (C) and (D).

$$D_{\text{ref}}: D_{\text{ref}}(\vec{x}_q) = D[\![\vec{x}_{\text{ref}}]\!] \tag{6}$$

Using the definitions in (3), (5) and (6), the explicit form of extension function $\mathcal{F}$ in (2) becomes

$$D = D[\![\vec{x}_q]\!] = \mathcal{F}(L_{100}(\vec{x}_q), L_{\text{ref}}(\vec{x}_q), D_{\text{ref}}(\vec{x}_q)) \tag{7}$$

For each query point, its dose is a function of distance to 100% isosurface, distance to reference point and dose on reference point (7).

*2.1.3  Dose extension by LUT*

We adopted a three-dimensional LUT to represent the extension function $\mathcal{F}$:

$$D = D[\![\cdot]\!] = \text{LUT}[\![L_{100}(\cdot), L_{\text{ref}}(\cdot), D_{\text{ref}}(\cdot)]\!] \tag{8}$$

where $\vec{x}_q$ is denoted by a dot (figure 2). The LUT is indexed by $L_{100}$ (X axis), $L_{\text{ref}}$ (Y axis) and $D_{\text{ref}}$ (Z axis). The ranges of three axes are -10cm–20cm, -15cm–20cm and 5%–120%, respectively. We refer to the unit cubic volume of LUT as "bin" in this paper, as opposed to "voxel" used for patient volume. The bin size are 2mm*2mm*5%. Position of a bin is defined by its center. The value stored in each bin of LUT is the value of dose in percentage of prescription dose. Output of dose extension—the dose in query point $D[\![\vec{x}_q]\!]$—is obtained by 3D table lookup (figure 2, rightwards arrow) with $L_{100}(\vec{x}_q)$, $L_{\text{ref}}(\vec{x}_q)$ and $D_{\text{ref}}(\vec{x}_q)$ as table indexes (8). In table lookup, the value is obtained by tri-linear interpolation.. Table 1 shows the complete steps of dose extension algorithm.

| Table 1. Pseudo-code of volumetric dose extension |
|---|
| **Algorithm 1.** Volumetric dose extension from 100% isosurface and reference surface |
| **Inputs:** Lookup table(LUT), 100% isosurface $\boldsymbol{S}_{100\%}$, reference surface $\boldsymbol{S}_{\text{ref}}$, dose distribution on reference surface $D[\![\boldsymbol{S}_{\text{ref}}]\!]$ and patient volume $\boldsymbol{V}$ |
| **Output:** Extended dose $D$ |
| **Steps:** <br> 1. For each query point $\vec{x}_q$ in $\boldsymbol{V}$ (do in parallel by vectorization) <br>   a. Calculate $L_{100}(\vec{x}_q)$ by equation (3) <br>   b. Locate reference point $\vec{x}_{\text{ref}}$ by equation (4) <br>   c. Calculate $L_{\text{ref}}(\vec{x}_q)$ by equation (5) <br>   d. Assign $D_{\text{ref}}(\vec{x}_q)$ by equation (6) <br>   e. Evaluate dose distribution $D[\![\vec{x}_q]\!]$ by 3D table lookup, with $L_{100}(\vec{x}_q)$, $L_{\text{ref}}(\vec{x}_q)$ and $D_{\text{ref}}(\vec{x}_q)$ as indexes. <br> 2. Output D as extended dose |

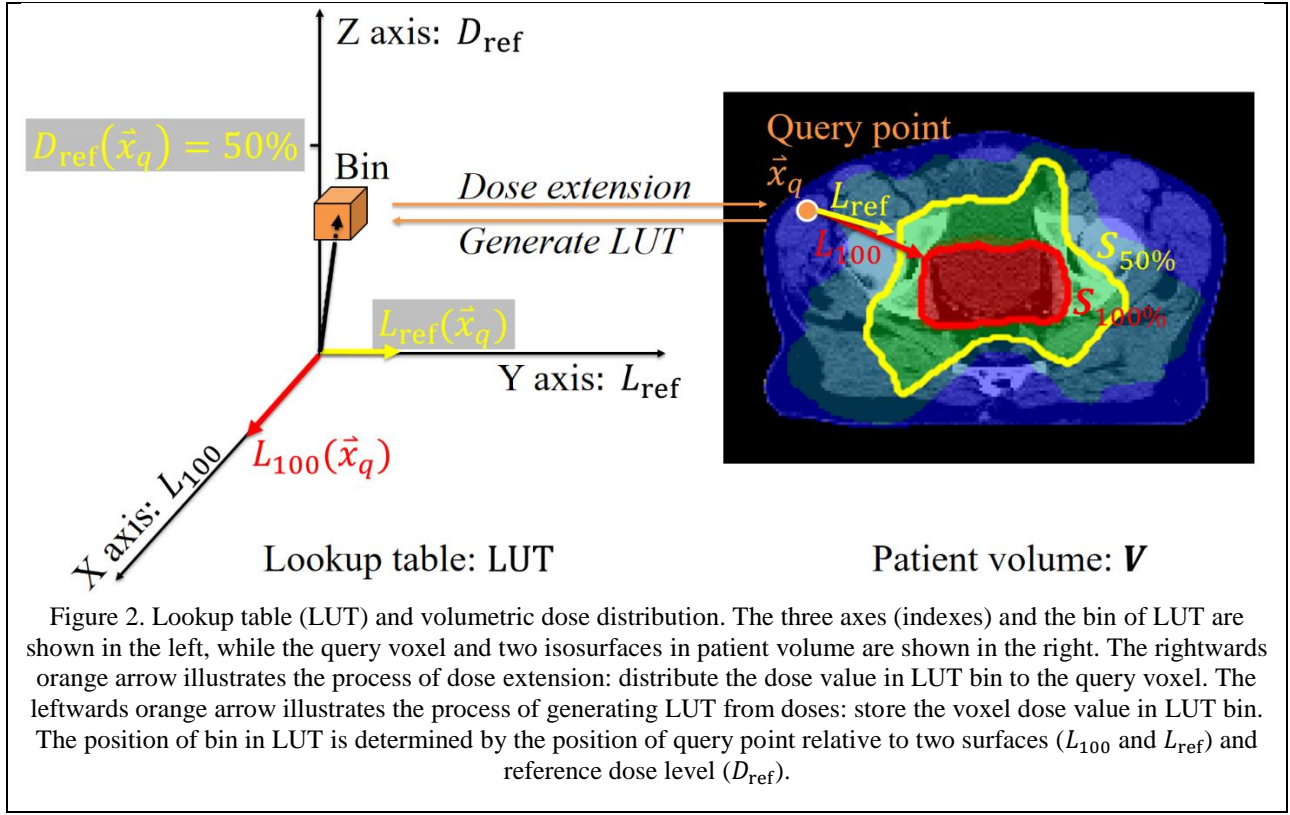

Figure 2. Lookup table (LUT) and volumetric dose distribution. The three axes (indexes) and the bin of LUT are shown in the left, while the query voxel and two isosurfaces in patient volume are shown in the right. The rightwards orange arrow illustrates the process of dose extension: distribute the dose value in LUT bin to the query voxel. The leftwards orange arrow illustrates the process of generating LUT from doses: store the voxel dose value in LUT bin. The position of bin in LUT is determined by the position of query point relative to two surfaces ($L_{100}$ and $L_{\text{ref}}$) and reference dose level ($D_{\text{ref}}$).

### 2.1.4 *Generate LUT from doses*

We used clinical dose distributions to generate the LUT. For individual 3D clinical dose distribution $d_k$, dose values of all voxels in $V_k$ are binned by $L_{100}$, $L_{\text{ref}}$ and $D_{\text{ref}}$ (figure 2, leftwards arrow). Three 3D tables—$N_k$, $T_k$ and $\sigma_k$—will be generated for dose $d_k$. Each bin of table $N_k$ counts the number of voxels stored in the bin. Table $T_k$ and table $\sigma_k$ keep the mean and standard deviation of the doses of the voxels stored in the bin. The generation of three tables from individual dose distribution is detailed in table 2.

| Table 2. Pseudo-code of LUT generation |
|---|
| **Algorithm 2.** Generate three LUTs from individual dose distribution |
| **Inputs:** Individual dose distribution $d_k$ and patient volume $V_k$ |
| **Output:** Table for number of voxels $N_k$, table for mean dose $T_k$ and table for standard deviation $\sigma_k$ |
| **Steps:** <br> 1. Locate $S_{100\%}$ and calculate distance map $L_{100}$ by equation (3) <br> 2. For each $z\%$ reference dose level (5%, 10%, … , 115%, 120%) <br>     a. Locate $S_{\text{ref}} = S_{t\%}$ and calculate distance map $L_{\text{ref}}$ by equation (5) <br>     b. For every voxel $\vec{x}$ in $V_k$ <br>         i. Store the voxel dose value $D[\![\vec{x}]\!]$ in the bin closest to $[L_{100}(\vec{x}), L_{\text{ref}}(\vec{x}), z\%]$ <br>         End for <br>     c. For each bin whose center $[x, y, z\%]$ in the range of $[-10\text{cm} \leq x \leq 20\text{cm},\ -15\text{cm} \leq y \leq 20\text{cm},\ z\%]$ <br>         i. Calculate the number of voxels, mean and standard deviation of the doses of the voxels that were stored in bin $[x, y, z\%]$ <br>         ii. Put the calculation results to the corresponding bin $[x, y, z\%]$ of $N_k$, $T_k$ and $\sigma_k$ <br>         End for <br>     End for |

The sets of three tables derived from individual dose distributions are combined by equation (9), in which all the operations are elementwise. The resulting table for number of voxels $N$, table for mean dose $T$ and table for standard deviation $\sigma$ capture the statistics of dose distributions in the training dataset.

$$\begin{cases} N = \sum_k N_k \\ T = \frac{1}{N} \sum_k N_k \cdot T_k \\ \sigma^2 = \frac{1}{N} \sum_k (N_k \cdot \sigma_k^2 + N_k \cdot T_k^2) - T^2 \end{cases} \quad (9)$$

The table $T$ is used for dose extension, and referred to as LUT by default in other sections, such as in equation (8). So $\text{LUT} \triangleq T$. The dose (in query voxel) evaluated by LUT reflects the mean dose of the voxels in the training dose distributions with the same geometrical configuration—distances to 100% isosurface and reference surface. We also obtain the table $\sigma$ that stores the standard deviation in training doses. We define $\text{LUT}^\sigma \triangleq \sigma$. By replacing the LUT in equation (8) with $\text{LUT}^\sigma$, we can estimate the uncertainty of extended dose $D^\sigma$ by equation (10)

$$D^\sigma = D^\sigma[\![\cdot]\!] = \text{LUT}^\sigma[\![L_{100}(\cdot),\ L_{\text{ref}}(\cdot),\ D_{\text{ref}}(\cdot)]\!] \quad (10)$$

## 2.2 Isodose tuning

### 2.2.1 Definition: dose tuning by isosurface dragging

Isodose tuning is to tune dose distribution by dragging an isosurface. The three steps are specified as follows. 1) A base dose $\dddot{D}^s$ (figure 4.a&c) is firstly optimized. The triple-dot accent notes that the dose distribution is actual and associated to an existing plan. The superscript $s$ ($s$ from *source*) notes that the dose is base dose, which would be tuned by isosurface dragging. 2) Then the base dose $\dddot{D}^s$ is presented to the planner (figure 4.c), who drags the isosurface of base dose (figure 4.d) to a desired position (figure 4.e). 3) Finally, isodose tuning is to obtain a new 3D dose distribution (figure 4.f, desired dose) whose isosurface is in the desired position. The ground truth (GT) for the desired dose is called as $\dddot{D}^t$ (figure 4.b). The triple-dot accent specifies that the GT should be actual: a plan that can deliver it must exist. The superscript $t$ ($t$ from *target* or *destination*) specifies that the dose is desired with isosurface in the desired position. "Desired" is the antonym of "base" in the context of this paper with different superscripts.

The task of isodose tuning can be achieved by an optimization approach, which initializes the optimization by the base dose, sets the objectives according to the desired isosurface position and obtains the actual desired dose $\ddot{D}^t$ by optimization. However, optimization is computation-intensive and time-consuming. To meet the need of interactive tuning, we apply the dose extension method in isodose tuning, as the table lookup operation is much faster than expensive iterative dose / derivative calculation. We also use a correction strategy to improve the accuracy of extended dose. The implementation is detailed in the next subsection.

*2.2.2   Implementation: dose extension with correction*

In the process of isosurface dragging, the $z\%$ isosurface of base dose $\ddot{D}^s$ (figure 4.d) was dragged to the desired position $S_{z\%}$, with 100% isosurface $S_{100\%}$ fixed (figure 4.e). The resulting dose of tuning action (isosurface dragging) is imagined in figure 4.f. $\ddot{D}^s$, $S_{z\%}$ and $S_{100\%}$ are inputs to isodose tuning algorithm, which obtains the output dose (figure 4.l) by extension with correction. The GT for the unknown tuned dose (figure 4.f) is the actual desired dose $\ddot{D}^t$ (figure 4.b), which is unknown (except for the positions of two isosurfaces) unless a plan optimization is carried out. We employ extended dose as approximate dose estimation and use a correction strategy[3, 15] to improve the accuracy. The correction strategy[3, 15] makes use of the information in the base plan to correct the approximate dose in the desired plan. The isodose tuning is implemented in four steps.

In the first step, we extend the desired dose from 100% isosurface and the desired position of $z\%$ isosurface (figure 4.h) by equation (11). The reference surface is isosurface, so $\ddot{D}^s[\![S_{z\%}]\!] \equiv z\%$ (notice the color of reference surface in figure 4.h) and $z\%$ is used in equation (11). The resulting extended dose is $\tilde{D}^t$ (figure 4.j). The tilde accent notes that the dose is an approximate, whose accent is different to that of actual dose. The approximate dose is obtained by dose extension, whose corresponding plan may not exist. So the "approximate" dose or "extended" dose is antonym of "actual" dose in the context, with different accents. The superscript $t$ indicates that the extended dose is calculated to approximate the desired dose, as the dose distribution on reference surface are extracted from the desired dose.

$$\tilde{D}^t = \tilde{D}^t[\![\cdot]\!] = \mathcal{F}(\cdot, S_{100\%}, S_{z\%}, z\%) \tag{11}$$

In the second step, we extend the base dose from $S_{100\%}$ and reference surface (figure 4.g) by equation (12). We set the desired $z\%$ isosurface $S_{z\%}$ as the reference surface, so the base doses distributed on reference surface $\ddot{D}^s[\![S_{z\%}]\!]$ are not uniform (notice the color of reference surface in figure 4.g). The resulting extended dose is $\tilde{D}^s$ (figure 4.i). The tilde accent notes that $\tilde{D}^s$ is an approximate dose obtained by dose extension, so $\tilde{D}^s \neq \ddot{D}^s$ (actual dose).

$$\tilde{D}^s = \tilde{D}^s[\![\cdot]\!] = \mathcal{F}(\cdot, S_{100\%}, S_{z\%}, \ddot{D}^s[\![S_{z\%}]\!]) \tag{12}$$

In the third step, we calculate the difference between actual dose and approximate (extended) dose in the base plan $s$ by equation (13). The difference $\Delta D^s$ is called as a correction map (figure 4.k).

$$\text{Correction: } \Delta D^s = \ddot{D}^s - \tilde{D}^s \tag{13}$$

In the final step, we apply the correction (figure 4.k) to the approximate (extended) desired dose (figure 4.j) by equation (14). The resulting dose is denoted by $\hat{D}^t$ (figure 4.l), where the hat accent notes that it is an extended dose with correction applied. $\hat{D}^t$ (figure 4.l) is the output of isodose tuning algorithm.

$$\hat{D}^t = \tilde{D}^t + \Delta D^s \tag{14}$$

To evaluate the accuracy of output $\hat{D}^t$ (extended dose with correction), we will compare it to GT—the actual desired dose $\ddot{D}^t$ (figure 4.b).

## 2.3 Experimental validation

We collected 32 clinical post-operative prostate volumetric modulated arc therapy (VMAT) plans from our institution. The PTV was prescribed by 25Gy in each plan. The PTV volume ranges from 200cc to 400cc, covering the post-operative prostate bed. The dose distributions in 27 training plans were used to build the LUTs (LUT and $\text{LUT}^\sigma$). We used the doses in five testing plans as GT to test the LUT-based extension (figure 3.a) and we simulated the isodose tuning process based on the first testing case (figure 4.b).

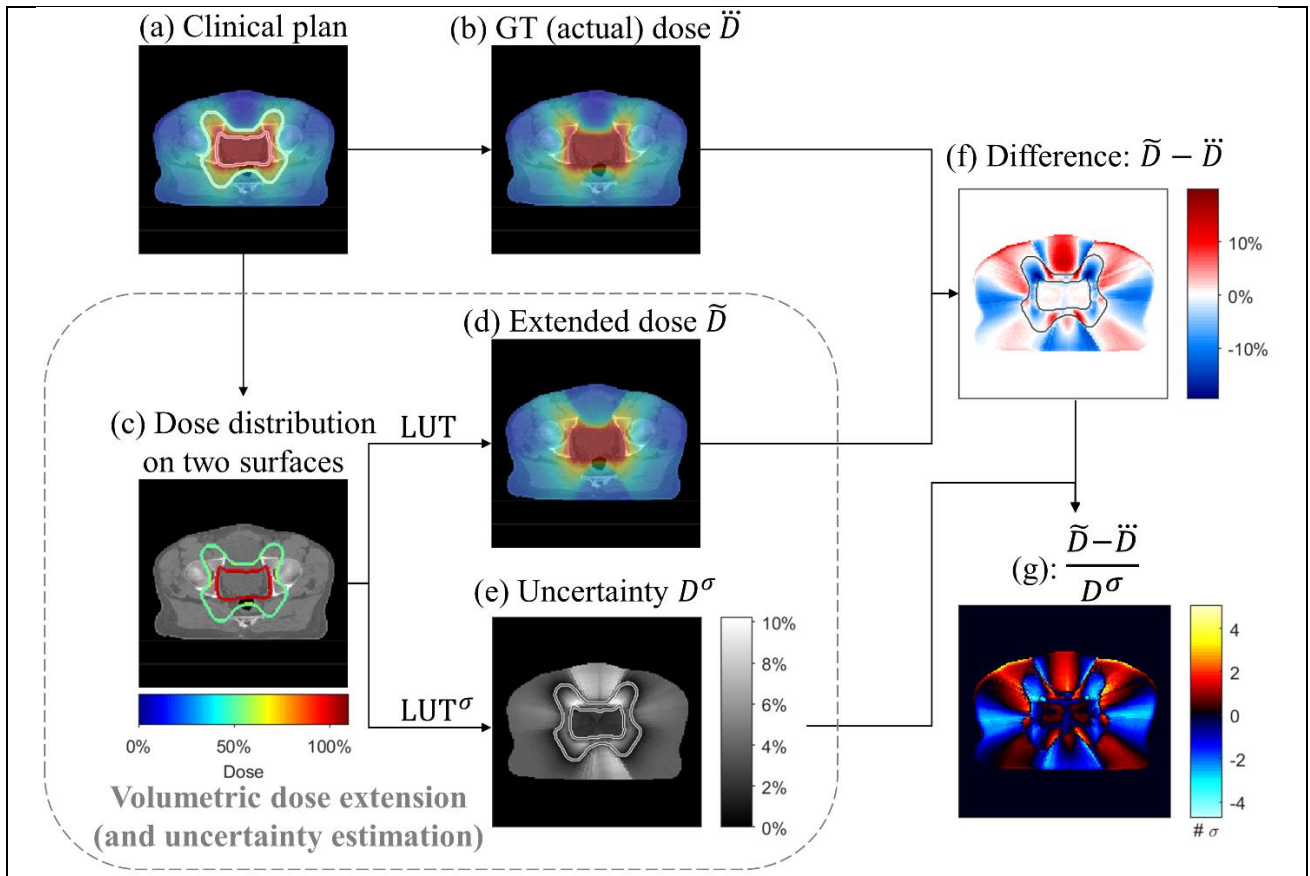

Figure 3. Experiment #1: Volumetric dose extension. The contents in the gray box represent the processes of volumetric dose extension and uncertainty estimation. The contents outside of box are for data preparation and result evaluation. (a) The clinical plan of the testing case. (b) The clinical (actual) dose $\ddot{D}$ is defined as GT dose. (c) Input of dose extension: dose distribution on two surfaces (surfaces colored by dose). (d) Extended dose distribution $\tilde{D}$. The colorbar for (a), (b), (c) and (d) is in the bottom-left corner. (e) Uncertainty of extended dose $D^\sigma$. (f) The difference between extended dose and GT. (g) The map of difference in number of $\sigma$. All the subfigures shown are from the first testing case.

The first experiment (figure 3) evaluated the accuracy of dose extension method. The doses in the five testing plans are referred to as $\ddot{D}$ (figure 3.b). We used LUT to extend the dose from its 100% and 50% isosurfaces (figure 3.c) to entire volume by equation (8). The resulting extended doses are called as $\tilde{D}$ (figure 3.d). The extended dose $\tilde{D}$ will be compared to GT dose—$\ddot{D}$—by dose volume histogram (DVH), gamma passing rate (GPR) and profiles. We also used $LUT^\sigma$ to calculate the uncertainty of extended dose (figure 3.e), and compare the uncertainty $D^\sigma$ to the discrepancy between extended dose $\tilde{D}$ and GT $\ddot{D}$ (figure 3.f).

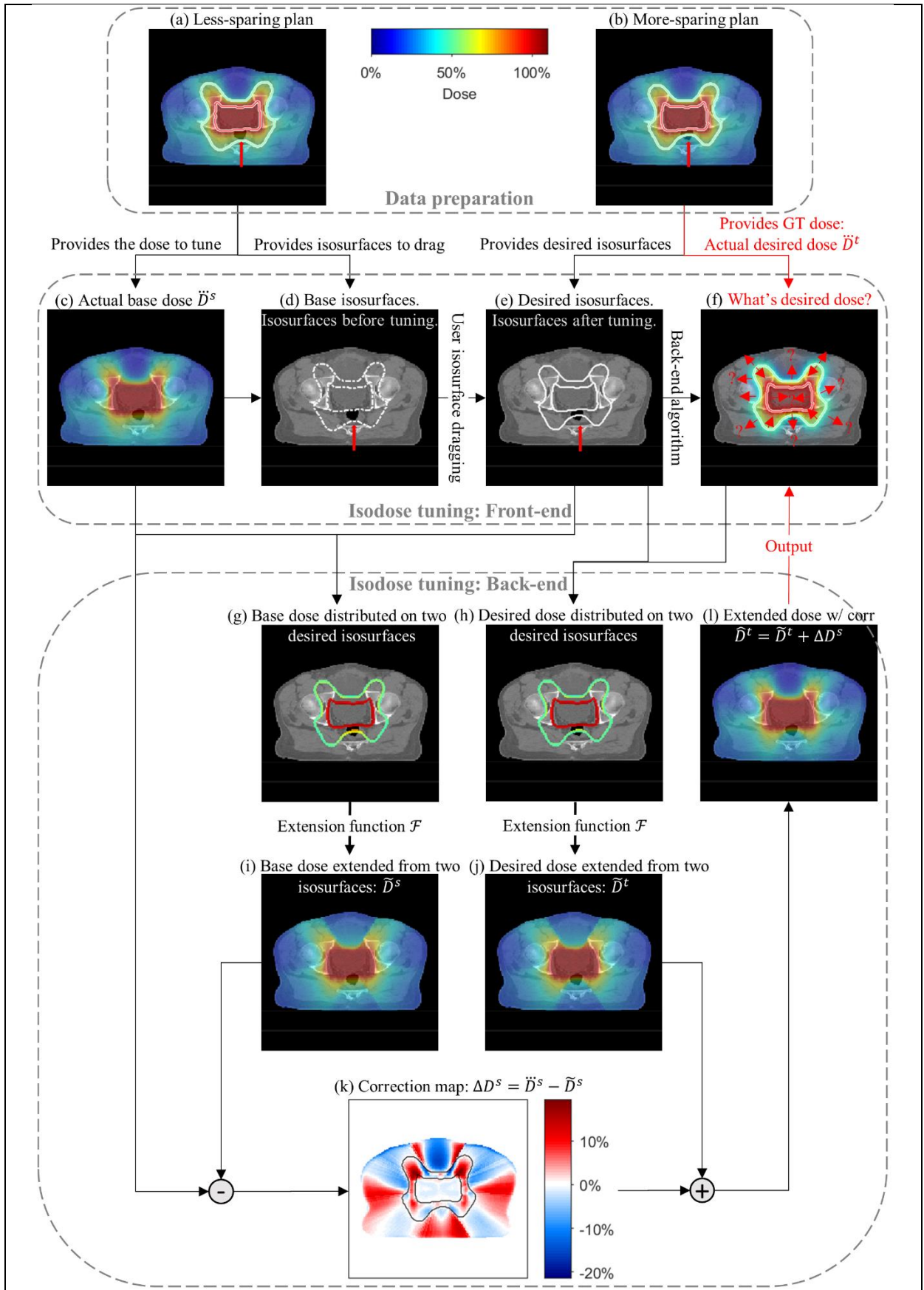

> Figure 4. Experiment #2: isodose tuning. The first block shows the process of preparing experimental data. The second block shows the front-end of isodose tuning and user interaction. User dragged the 50% isosurface in rectum region closer to target, so more rectum is spared (red arrows in subfigure (d)&(e)). The third block shows the back-end algorithm of isodose tuning. The colorbar for all subfigures except for k is shown in the top block. Subfigure (b) provides the GT for the unknown tuning result (f) and subfigure (l) is the output of isodose tuning algorithm. As the isodose tuning was simulated based on the first testing case, subfigure (b) and figure 3.a, subfigure (h) and figure 3.c, subfigure (j) and figure 3.d are the same.

The second experiment evaluated the accuracy of isodose tuning (figure 4). To simulate the scenario of isodose tuning, we re-optimized the clinical plan in the first testing case (testing case #1), and obtained a dose distribution with less rectum-sparing (figure 4.a, red arrow) than the dose in clinical plan (figure 4.b, red arrow) while keeping 100% isosurface the same. The less-sparing plan was defined as the base plan and the more-sparing plan (clinical) was defined as the desired plan. The doses in the base plan and desired plan—$\ddot{D}^s$ (figure 4.c) and $\ddot{D}^t$ (figure 4.b)—are actual doses as they are associated from existing plans. Volumetric dose distribution of $\ddot{D}^s$ (figure 4.c) and the positions of 100% and 50% isosurfaces of $\ddot{D}^t$ (figure 4.e) are the inputs to isodose tuning. Volumetric dose distribution of $\ddot{D}^t$ (figure 4.b) is the GT for tuning result (figure 4.f). We performed the four-step isodose tuning algorithm in Section 2.2.2, and compared the output—extended dose with correction $\hat{D}^t$ (figure 4.l)—to the GT $\ddot{D}^t$ (figure 4.b) in terms of DVH, GPR and profiles. Note that the GT dose $\ddot{D}$ (figure 3.b) in experiment #1 is the also the GT for desired dose $\ddot{D}^t$ (figure 4.b) in this isodose tuning experiment, so the results in experiment #1 reflect the accuracy of dose extension itself in isodose tuning with no correction applied.

## 3 Results
### 3.1 LUT

The LUT and LUT$^\sigma$ calculated from 27 clinical post-operative prostate VMAT plans are displayed in figure 5. The LUT storing mean values are displayed in figure 5.a. We can observe the dose decreases as $L_{100}$ increases. The data entries only distribute in the region where $L_{100} > L_{\text{ref}}$, because $S_{50\%}$ is at the outside (positive direction) of $S_{100\%}$. The entries whose $L_{100} \leq 0$ are inside $S_{100\%}$, so their dose should be 100%, which is consistent with the result in figure 5.a. To further understand the LUT, consider a rotational-symmetric geometry where the 100% and 50% isodose lines are two concentric circles on a plane. Denote $b$ as the spacing between two circles (distance from $S_{100\%}$ to $S_{50\%}$), $x$ as distance to inner circle ($L_{100}$) and $y$ as distance to outer circle ($L_{\text{ref}}$). For any query point on the plane, $y = x - b$. Dose falloff profiles can be extracted from LUT along parallel lines $y = x - b$ with different $b$ values. The dose falls off faster along the line $y = x - b$ with smaller $b$, which corresponds to tighter $S_{50\%}$. The LUT$^\sigma$ storing standard deviations are displayed in figure 5.b. As the dose value on the two isosurfaces is fixed, the query point whose $L_{100} = 0$ or $L_{\text{ref}} = 0$ should have zero uncertainty. This is consistent with the result in figure 5.b that standard deviation is zero when $x = 0$ or $y = 0$. We also observe that the data entries closer to two axes have smaller standard deviation σ (uncertainty), which means that the query points closer to isosurface will have smaller dose uncertainty. In the other region, the order of magnitude is about 5%. If we assume the voxel doses stored in one LUT bin (figure 2, leftwards arrow) follow Gaussian distribution, then the uncertainty of extended dose is ±5% in ±1σ interval. The data entries with the highest uncertainty (~8%) are on the line $y = x - b$ with a very small value of $b$. This means that if the reference 50% isosurface is very close to the 100% isosurface (very tight along some direction), the dose in query points will be most uncertain.

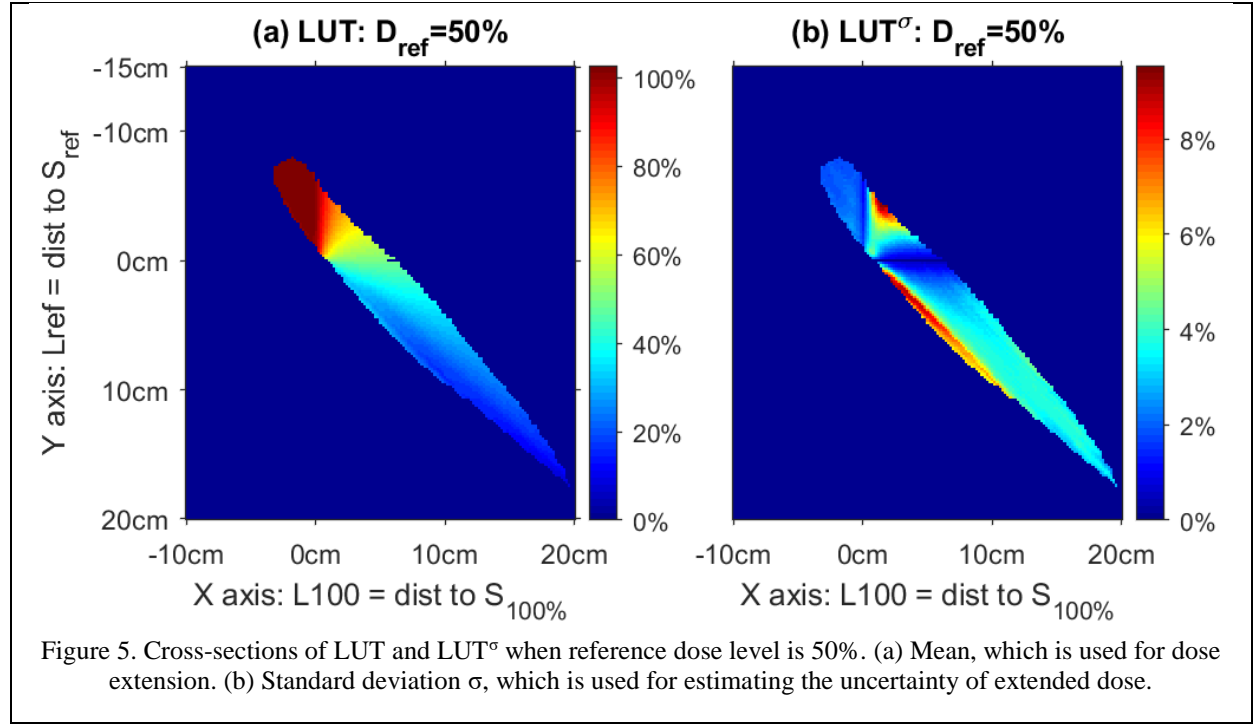

Figure 5. Cross-sections of LUT and LUT$^\sigma$ when reference dose level is 50%. (a) Mean, which is used for dose extension. (b) Standard deviation σ, which is used for estimating the uncertainty of extended dose.

## 3.2 Accuracy of dose extension

To evaluate the accuracy of volumetric dose extension (table 1), we compared the extended dose $\widetilde{D}$ (figure 3.d) to GT dose $\ddot{D}$ (figure 3.b) in five testing cases. As shown in figure 3.f, the difference $\widetilde{D} - \ddot{D}$ is around 10%. To quantify the difference, we calculated the GPR and maximum/mean dose difference of five testing cases in table 3. The mean differences are 3-5% while the extreme differences can be 10%-16%.

Table 3. Accuracy of dose extension method. GPRs of the extended doses $\widetilde{D}$ in five testing cases are listed in the first block. Regions whose dose is lower than 10% are excluded. The absolute values of difference in the region between 20% and 50% isosurfaces, the region between 50% and 80% isosurfaces, and the region between 80% and 100% isosurfaces are listed in the second block. Top 2% outliers were removed before maximum value was evaluated.

| Criterion | 2mm/2% | 3mm/3% | 5mm/5% |
|---|---|---|---|
| GPR($\widetilde{D}$) | 63.0% ± 4.4% | 77.2% ± 3.5% | 92.0% ± 3.0% |
| Region | 20%~50% | 50%~80% | 80%~100% |
| Max($|\widetilde{D} - \ddot{D}|$) | 10.6% ± 1.3% | 11.9% ± 2.8% | 16.7% ± 2.7% |
| Mean($|\widetilde{D} - \ddot{D}|$) | 3.1% ± 0.5% | 2.4% ± 0.3% | 4.5% ± 0.6% |

The difference between extended dose and GT originates from the uncertainty of LUT. Therefore, we used LUT$^\sigma$ to calculate the uncertainty of extended dose. The uncertainty map $D^\sigma$ (figure 3.e) shows the standard deviation of extended dose per voxel. We divided the difference map $\widetilde{D} - \ddot{D}$ (figure 3.f) by $D^\sigma$ (figure 3.e), and obtained a map of different in the unit of standard deviation $\sigma$ (figure 3.g). The differences of most voxels are visually in the three $\sigma$ interval (±3$\sigma$). We calculated the histogram of dose difference measured by the number of $\sigma$ for five testing cases (figure 6). The difference of 62.2%±8.5% voxels in patient volume is in the one $\sigma$ interval, 90.5%±3.1% in two $\sigma$ interval and 97.5%±0.7% in three $\sigma$ interval.

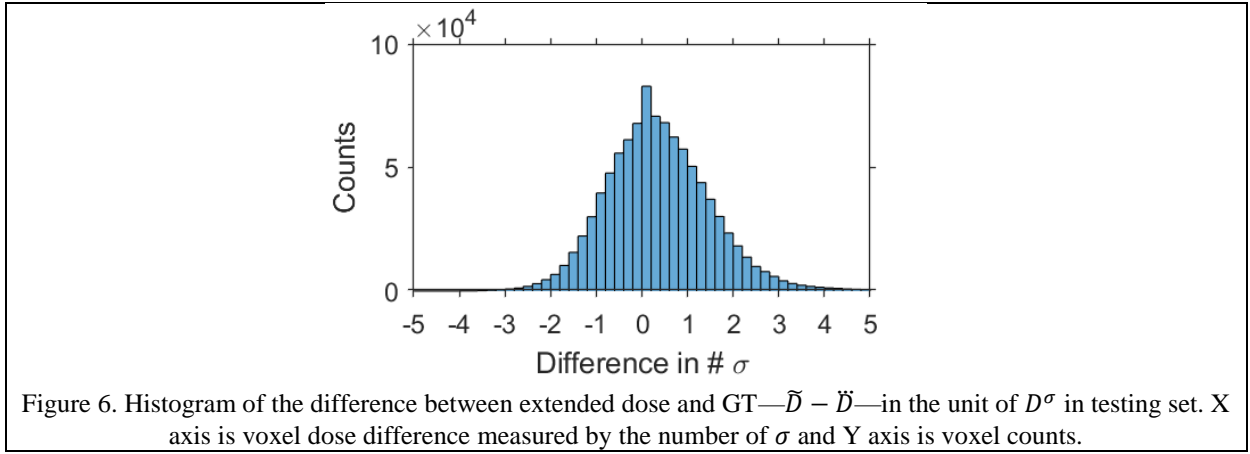

Figure 6. Histogram of the difference between extended dose and GT—$\tilde{D} - \ddot{D}$—in the unit of $D^\sigma$ in testing set. X axis is voxel dose difference measured by the number of $\sigma$ and Y axis is voxel counts.

We compared the extended dose $\tilde{D}$ to GT dose $\ddot{D}$ in DVH for testing case #1 in figure 7, which matches within the uncertainty level reported in table 3. The rectum matches the best as it is a small volume close to the reference surface (figure 3.c), the uncertainty is small (Section 3.1) and thus the dose difference is small. The profiles of extended dose and GT are compared in figure 10 for testing case #1. The 10% difference is reflected by the positions of blue ($\ddot{D}$) and blue dotted ($\tilde{D}$) profiles.

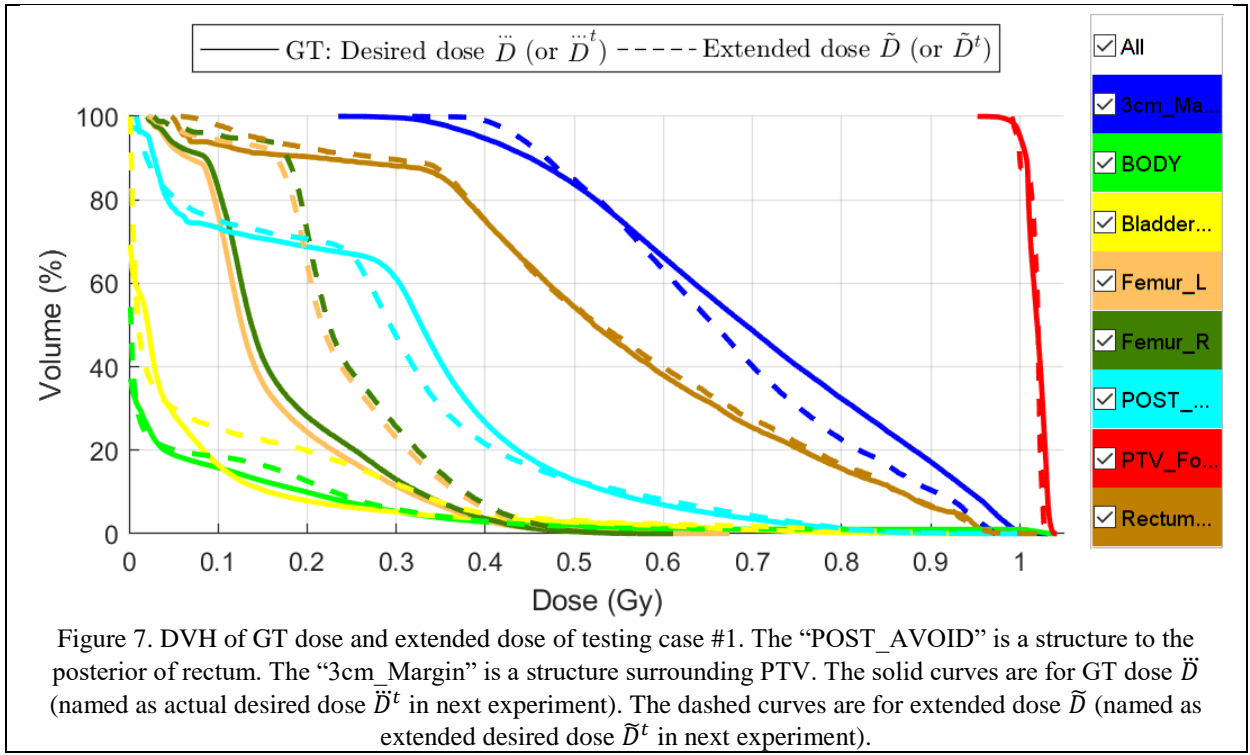

Figure 7. DVH of GT dose and extended dose of testing case #1. The "POST_AVOID" is a structure to the posterior of rectum. The "3cm_Margin" is a structure surrounding PTV. The solid curves are for GT dose $\ddot{D}$ (named as actual desired dose $\ddot{D}^t$ in next experiment). The dashed curves are for extended dose $\tilde{D}$ (named as extended desired dose $\tilde{D}^t$ in next experiment).

### 3.3 Accuracy of isodose tuning: dose extension with correction

The accuracy of isodose tuning algorithm is evaluated by the similarity between output $\hat{D}^t$ (extended dose with correction, figure 4.l) and GT $\ddot{D}^t$ (figure 4.b). Their difference, $\hat{D}^t - \ddot{D}^t$, is plotted in figure 8. The dose difference in most region is ±2% in visual inspection.

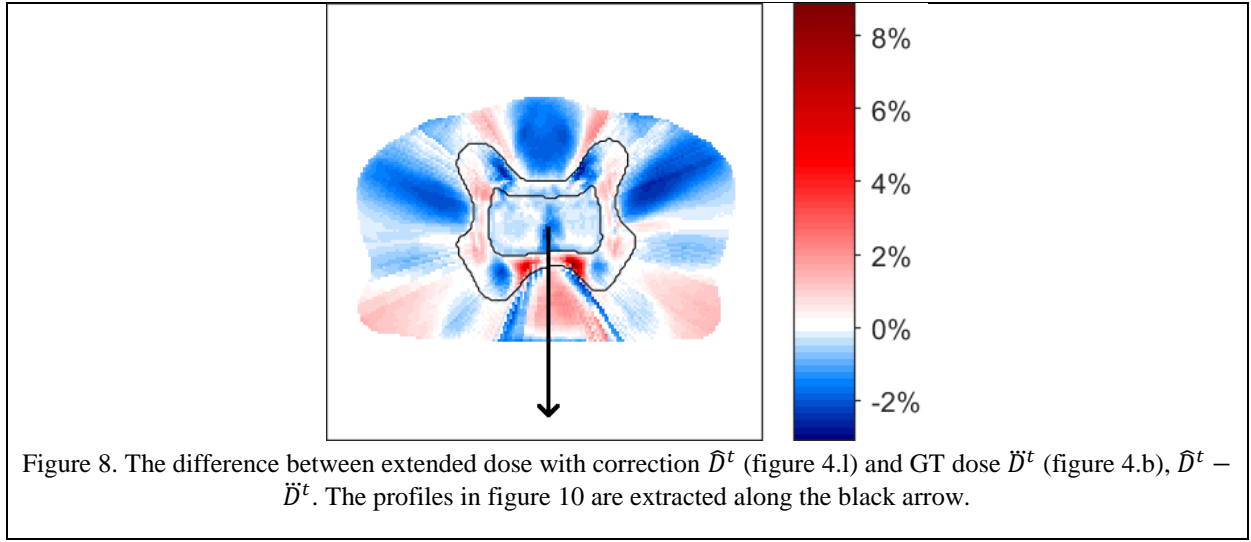

Figure 8. The difference between extended dose with correction $\hat{D}^t$ (figure 4.l) and GT dose $\ddot{D}^t$ (figure 4.b), $\hat{D}^t - \ddot{D}^t$. The profiles in figure 10 are extracted along the black arrow.

We quantify the difference by GPR and maximum/mean dose difference in table 4. The extended dose with correction $\hat{D}^t$ (figure 4.l) is very close to the GT (actual dose, figure 4.b) with 3mm/3% GPR higher than 97% and mean dose difference less than 1% in all three regions. Compared to the raw extended dose without correction $\tilde{D}^t$ (figure 4.j), the correction step significantly improve the accuracy. GPR (3mm/3%) is increase by 20.9%, maximum dose difference between 80% and 100% isosurfaces is decreased by 10% and mean dose difference decrease by 3.1%.

Table 4. Accuracy of isodose tuning. GPRs of the output—extended dose with correction $\hat{D}^t$ (figure 4.l)—are listed in the first block. The absolute values of difference in three regions are listed in the second block. The computation settings are the same as table 3. The GPRs and dose differences of extended dose (without correction) $\tilde{D}^t$ (figure 4.j) are listed in parentheses to the right side to demonstrate the effect of correction strategy.

| Criterion | 2mm/2% | 3mm/3% | 5mm/5% |
|---|---|---|---|
| GPR($\hat{D}^t$) | 88.5% (60.8%) | 97.2% (76.3%) | 99.9% (92.4%) |
| Region | 20%~50% | 50%~80% | 80%~100% |
| Max($|\hat{D}^t - \ddot{D}^t|$) | 3.6% (9.0%) | 4.2% (8.6%) | 3.9% (13.9%) |
| Mean($|\hat{D}^t - \ddot{D}^t|$) | 0.9% (2.7%) | 0.7% (2.0%) | 0.9% (4.0%) |

We plot the DVH curves for GT dose, output and base dose in figure 9. The dash-dotted curves are for the dose before tuning action $\ddot{D}^s$, and the solid curves are the GT of the dose after tuning action $\ddot{D}^t$. The output of our isodose tuning method $\hat{D}^t$ (dash-dotted) is very close to GT in DVH curves, especially in rectum and post-avoid structure, for whose sparing the isosurface was dragged (figure 4.d&e).

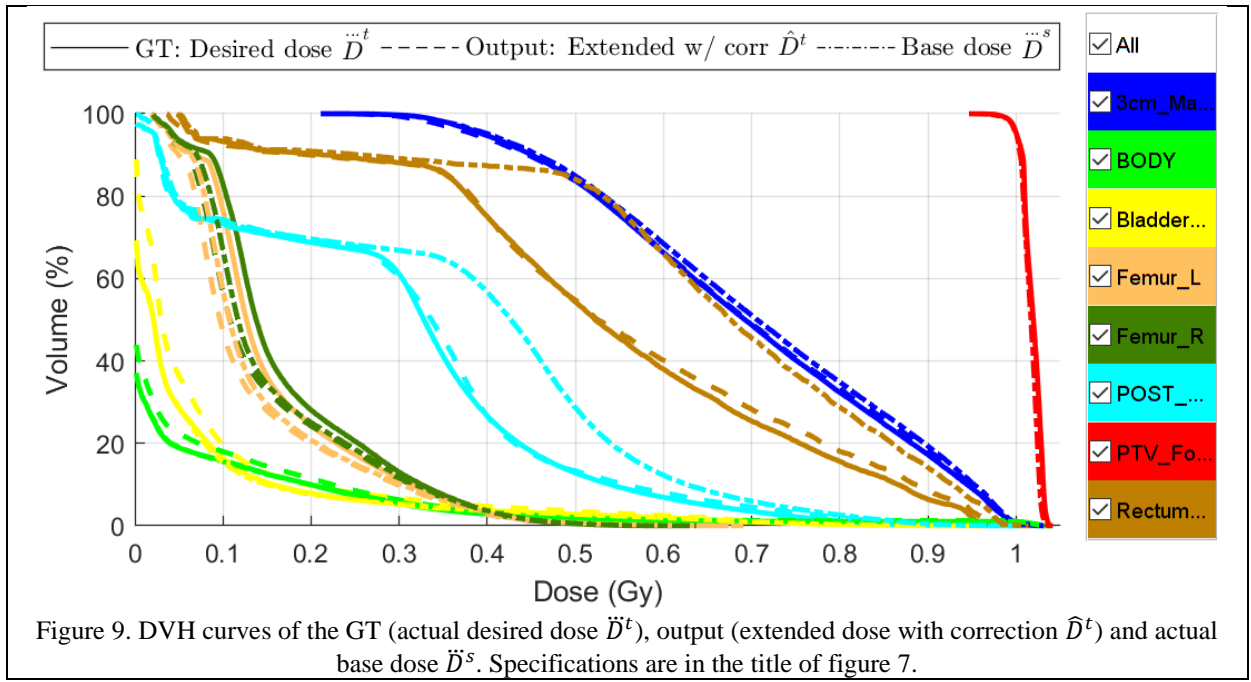

Figure 9. DVH curves of the GT (actual desired dose $\ddot{D}^t$), output (extended dose with correction $\hat{D}^t$) and actual base dose $\ddot{D}^s$. Specifications are in the title of figure 7.

We plotted the line profiles of doses to illustrate the correction strategy. The correction map (green curve subtracted by green dotted curve) is obtained by evaluating the difference between actual dose $\ddot{D}^s$ and approximate dose $\tilde{D}^s$ (extended dose) in base plan $s$ (13). Then the correction is applied to extended desired dose $\tilde{D}^t$ (blue-dotted curve) by equation (14). The resulting output $\hat{D}^t$ (red curve) is close to GT $\ddot{D}^t$ (blue curve).

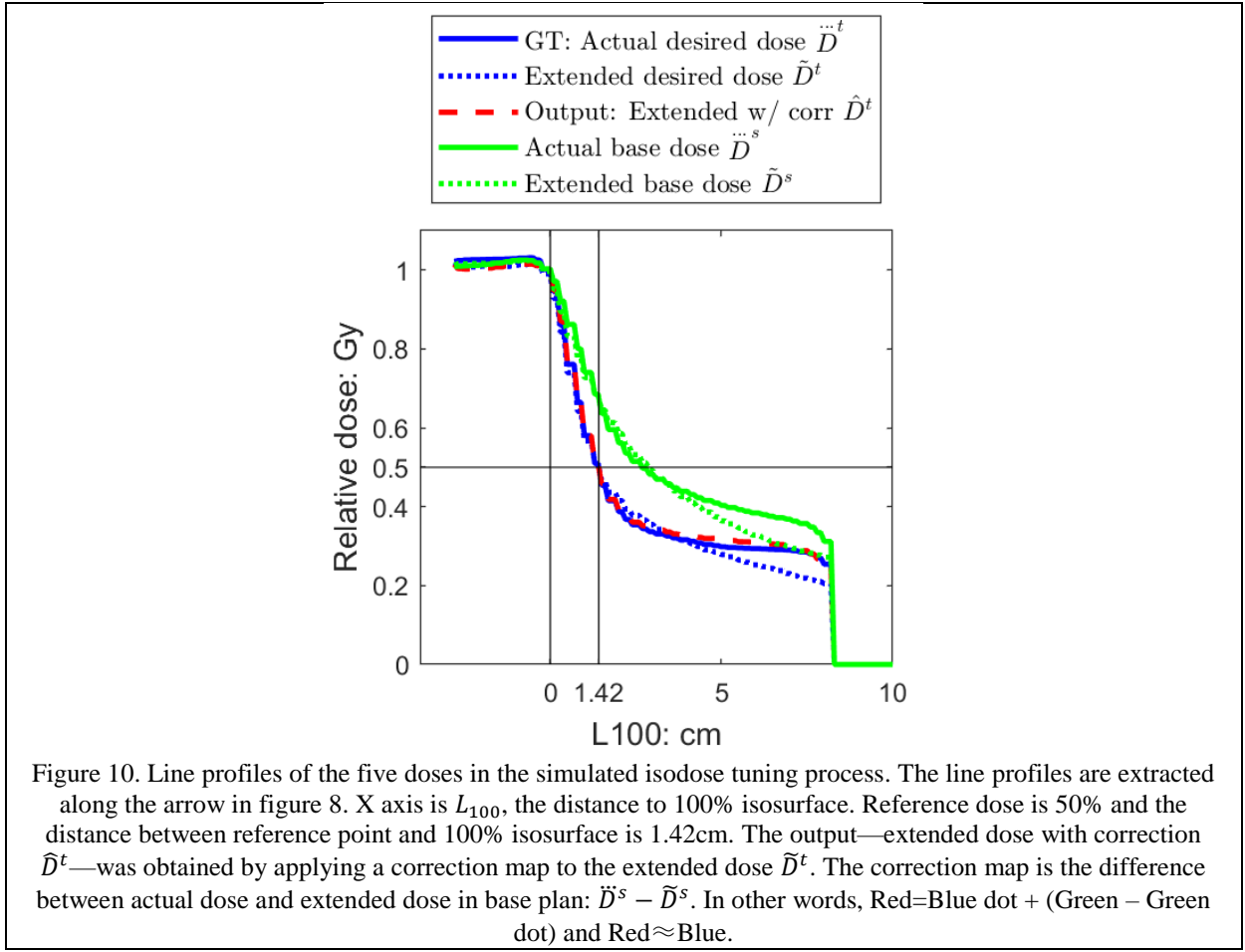

Figure 10. Line profiles of the five doses in the simulated isodose tuning process. The line profiles are extracted along the arrow in figure 8. X axis is $L_{100}$, the distance to 100% isosurface. Reference dose is 50% and the distance between reference point and 100% isosurface is 1.42cm. The output—extended dose with correction $\hat{D}^t$—was obtained by applying a correction map to the extended dose $\tilde{D}^t$. The correction map is the difference between actual dose and extended dose in base plan: $\dddot{D}^s - \tilde{D}^s$. In other words, Red=Blue dot + (Green – Green dot) and Red≈Blue.

The dose extension and isodose tuning algorithm are implemented in Matlab® R2017b (The Mathworks, Inc) without parallelization and evaluated with a Windows 10 workstation with 8-core 3.60 GHz CPU. The computation time of dose extension is less than two seconds, during which the calculation of $L_{100}$, $L_{\text{ref}}$ and $D_{\text{ref}}$ takes ~1.5 seconds and the extension operation (table lookup) takes less than 0.1 seconds. The computation time of isodose tuning is also less than two seconds. As the extensions of base dose and desired dose are from the same surfaces (equation (11) and (12)), the calculation of $L_{100}$, $L_{\text{ref}}$ and $D_{\text{ref}}$ only happens once. The table lookup operations happens twice and take less than 0.2 seconds. Obtaining (13) and applying (14) correction map are just matrix arithmetic operations, which are negligible compared with other parts. Therefore, the response time of isodose tuning algorithm is fast enough and it can be used for interactive tuning.

## 4   Discussion

We presented a novel dose extension method that can extend dose distribution from two isosurfaces to the entire patient volume with distances as extension variables. We defined query point–to-surface distances for 100% isosurface and reference surface respectively, represented the extension function by a LUT and validated the extension method for post-operative prostate VMAT dose distributions. This paper initiated the research on dose extension—generating dose distribution from two isosurfaces—and set a general framework for it. More researches on dose extension can be conducted regarding validation in more treatment sites, new ways of defining point-to-surface distances, locating reference point and formulating extension function. More geometry-related information may be used as input to lower the uncertainty of extended dose distribution.

We demonstrated the application of dose extension in the plan isodose tuning. We adopted it as an approximate dose estimation and used a correction strategy[3, 15] to improve the accuracy. Traditional interactive dose tuning[16] relied on optimization to tune (re-optimize from) the dose in base plan. The operation of tuning is in the same way as in conventional treatment planning, such as setting dose constraints and tuning weights of each ROI. A simplified dose calculation engine[16] was used in optimization to satisfy the speed needed for interactive tuning. The novel interactive tuning method proposed in this paper—dose extension with correction—utilized the base plan through a correction strategy. It is operated by isosurface dragging and based on dose extension. As the computation cost (distance transform and table lookup) of dose extension is marginal, real-time interactive isodose tuning is feasible. To simulate the isodose tuning process, we used the isosurface in an existing plan as the results of tuning action of users. We will explore methods and graphic user interface that can guide the isosurface dragging for users in the future.

The dose extension and isodose tuning generate dose distributions without associated plans. To deliver the dose output by them, we can perform dose mimicking optimization[16], which were also used by DL-based dose prediction methods[9, 14, 17] to realize their predicted doses. Recently we proposed a DL-based fluence map prediction method[18] that directly predicts fluence maps for desired dose without optimization. We will try to combine the isodose tuning with fluence map prediction in the future, so a fluence-based plan can be generated interactively in response to isosurface dragging.

## 5    Conclusion

We proposed a novel dose extension method, which extends dose distribution from two isosurfaces to entire patient volume. The dose as a function of distances to isosurfaces was stored in a LUT. The dose extension is implemented by the computation-efficient table lookup. We demonstrated the application of dose extension with isodose tuning in radiotherapy planning. In response to tuning action (isosurface dragging), the output dose was obtained by dose extension with correction, which is accurate and efficient.

**References**
1.	Lu W, Chen M. Fluence-convolution broad-beam (FCBB) dose calculation. *Physics in Medicine & Biology*. 2010;55(23):7211.
2.	Chen Q, Chen M, Lu W. Ultrafast convolution/superposition using tabulated and exponential kernels on GPU. *Medical physics*. 2011;38(3):1150-1161.
3.	Lu W. A non-voxel-based broad-beam (NVBB) framework for IMRT treatment planning. *Physics in Medicine & Biology*. 2010;55(23):7175.
4.	Yuan L, Ge Y, Lee WR, Yin FF, Kirkpatrick JP, Wu QJ. Quantitative analysis of the factors which affect the interpatient organ-at-risk dose sparing variation in IMRT plans. *Medical physics*. 2012;39(11):6868-6878.
5.	Appenzoller LM, Michalski JM, Thorstad WL, Mutic S, Moore KL. Predicting dose-volume histograms for organs-at-risk in IMRT planning. *Medical physics*. 2012;39(12):7446-7461.
6.	Ahmed S, Nelms B, Gintz D, et al. A method for a priori estimation of best feasible DVH for organs-at-risk: Validation for head and neck VMAT planning. *Medical physics*. 2017;44(10):5486-5497.
7.	Nguyen D, Long T, Jia X, et al. A feasibility study for predicting optimal radiation therapy dose distributions of prostate cancer patients from patient anatomy using deep learning. *Scientific reports*. 2019;9(1):1-10.

8.        Nguyen D, Jia X, Sher D, et al. 3D radiotherapy dose prediction on head and neck cancer patients with a hierarchically densely connected U-net deep learning architecture. *Physics in Medicine & Biology*. 2019;64(6):065020.
9.        Babier A, Mahmood R, McNiven AL, Diamant A, Chan TCY. Knowledge-based automated planning with three-dimensional generative adversarial networks. *Medical Physics*. 2019;47(2)doi:10.1002/mp.13896
10.       Barragán-Montero AM, Nguyen D, Lu W, et al. Three-dimensional dose prediction for lung IMRT patients with deep neural networks: robust learning from heterogeneous beam configurations. *Medical physics*. 2019;46(8):3679-3691.
11.       Barragán-Montero AM, Thomas M, Defraene G, et al. Deep learning dose prediction for IMRT of esophageal cancer: The effect of data quality and quantity on model performance. *Physica Medica*. 2021;83:52-63.
12.       Nguyen D, Barkousaraie AS, Shen C, Jia X, Jiang S. Generating Pareto Optimal Dose Distributions for Radiation Therapy Treatment Planning. Springer International Publishing; 2019:59-67.
13.       Ma J, Bai T, Nguyen D, et al. Individualized 3D dose distribution prediction using deep learning. Springer; 2019:110-118.
14.       Nilsson V, Gruselius H, Zhang T, De Kerf G, Claessens M. Probabilistic dose prediction using mixture density networks for automated radiation therapy treatment planning. *Physics in Medicine & Biology*. 2021;66(5):055003.
15.       Siebers JV, Lauterbach M, Tong S, Wu Q, Mohan R. Reducing dose calculation time for accurate iterative IMRT planning. *Medical physics*. 2002;29(2):231-237.
16.       Otto K. Real-time interactive treatment planning. *Physics in Medicine & Biology*. 2014;59(17):4845.
17.       McIntosh C, Welch M, McNiven A, Jaffray DA, Purdie TG. Fully automated treatment planning for head and neck radiotherapy using a voxel-based dose prediction and dose mimicking method. *Physics in Medicine & Biology*. 2017;62(15):5926.
18.       Ma L, Chen M, Gu X, Lu W. Deep learning-based inverse mapping for fluence map prediction. *Physics in Medicine & Biology*. 2020;65(23):235035.